\documentstyle[preprint,prc,aps,epsf]{revtex}
\begin{document}
\draft
\title
{Approximate  Coulomb Distortion Effects in $(e,e'p)$ Reactions.}
\author{ K.S. Kim and  L.E. Wright}
\address{Institute of Nuclear and Particle Physics,
Department of Physics and Astronomy, Ohio University, Athens, 
Ohio 45701}

\maketitle
\begin{abstract}
In this paper we apply a well-tested approximation of electron  
Coulomb distortion effects to the exclusive reaction $(e,e'p)$ in  
the quasielastic region.  
We compare the approximate treatment of Coulomb distortion effects 
to the exact Distorted Wave Born Approximation evaluated by means  
of partial wave analysis to gauge  the quality of our approximate 
treatment.  
We show that the  approximate M${\o}$ller potential has a
plane-wave-like structure and hence permits the separation of the cross section 
into five terms which depend on bi-linear products of transforms of 
the transition four current elements.  
These transforms reduce to Fourier transforms when Coulomb 
distortion is not present, but become modified with the inclusion of 
Coulomb distortion.  
We investigate the application of the approximate formalism to a 
model of $^{208}Pb(e,e'p)$ using Dirac-Hartree single particle 
wavefunctions for the ground state and relativistic optical model 
wavefunctions for the continuum proton.  
We show that it is still possible to extract, albeit with some 
approximation, the various structure functions from the
experimentally measured data even for heavy nuclei.
\end{abstract}
\narrowtext

\section{Introduction}
Medium and high energy electron scattering has long been 
acknowledged as a useful tool in the investigation of nuclear 
structure and nuclear properties, especially in the quasielastic 
region.  
In the plane wave Born approximation (PWBA), where electrons are 
described by Dirac plane waves, the cross section for the exclusive 
reaction $(e,e'p)$ on nuclei can be written simply as
\begin{eqnarray}
{\frac {d^{3}{\sigma}} {dE_{f}d{\Omega}_{f}d{\Omega}_{P}}}
&=&{\frac {PE_{P}} {(2{\pi})^{3}}}{\sigma}_{M}[v_L R_{L}+ v_T R_{T}+
{\cos {2{\phi}_{P}}} v_{TT} R_{TT} \nonumber \\
&+&{\cos {{\phi}_{P}}} v_{LT} R_{LT}
+h {\sin {{\phi}_{P}}} v_{LT^{\prime}}R_{LT^{\prime}}] \label{crosep}
\end{eqnarray}
where $q_\mu ^2 = \omega^2-q^2$ is the four-momentum transfer,
$\sigma _{M}$ is the Mott cross section given by
$\sigma_{M} = \left(\frac{\alpha }{2E} \right)^2  \frac{\cos^2
\frac{\theta}{2}}{\sin^4 \frac{\theta}{2}}$, and $R_L$, $R_T$, 
$R_{TT}$, $R_{LT}$, and $R_{LT'}$ are the longitudinal, transverse, 
transverse-transverse, longitudinal-transverse, and polarized 
longitudinal-transverse structure functions which depend only on 
the momentum transfer $q$ and the energy transfer $\omega$.  The functions $v$ only
depend on electron kinematics and are given by
\begin{eqnarray}
v_L&=&{\frac {q^{4}_{\mu}}{q^{4}}} \nonumber \\
v_T&=&{\tan^{2} {\frac {{\theta}_{e}} 2}}-{\frac
{q_{\mu}^{2}} {2q^{2}}} \nonumber \\
v_{TT}&=&-{\frac {q_{\mu}^{2}} {2q^{2}}} \nonumber \\
v_{LT}&=&-{\frac {q_{\mu}^{2}} {q^{2}}}({\tan^{2} {\frac {{\theta}_{e}} 2}}
-{\frac {q_{\mu}^{2}} {q^{2}}})^{1/2}  \nonumber\\
v_{LT^{\prime}}&=&- {\frac {q_{\mu}^{2}} {q^{2}}}
{\tan {\frac {{\theta}_{e}} 2}} \label{elecst} 
\end{eqnarray}

Choosing the momentum transfer ${\bf q}$ to define the 
${\hat {\bf z}}$ axis and using the azimuthal symmetry of the spatial part 
of the M${\o}$ller potential permits the extraction of the explicit 
dependence on the azimuthal angle ${\phi}_{P}$ of the outgoing 
proton as measured with respect to the electron scattering plane. More
specifically, we define $\hat {\bf y} = \hat {\bf p}_i \times \hat {\bf p}_f$.  
These structure functions are defined as:
\begin{eqnarray}
R_{L}&=&{\frac {q^{4}} {q_{\mu}^{4}}}W_{00} \nonumber \\
R_{T}&=&W_{11}+W_{22} \nonumber \\
{\cos {2{\phi}_{P}}}R_{TT}&=&W_{11}-W_{22} \nonumber \\
{\cos {{\phi}_{P}}}R_{LT}&=&-{\frac {q^{2}} {q_{\mu}^{2}}}
(W_{01}+W_{10})  \nonumber\\
{\sin {{\phi}_{P}}}R_{LT^{\prime}}&=&-i{\frac {q^{2}} {q_{\mu}^{2}}}
(W_{02}+W_{20}) \label{structure}
\end{eqnarray}
where the nuclear tensors $W_{{\mu}{\nu}}$ are given in terms of a sum over $s_p$ and $\mu_p$
the final and initial spin projections of the nucleon,
\begin{equation}
W_{{\mu}{\nu}}=\sum_{s_{p}{\mu}_{b}}N^{\ast}_{\mu}N_{\nu}.
\label{nclten}
\end{equation}
and we have suppressed the spin labels for clarity.
The quantities $N^{\mu}$ are the Fourier transform of the nucleon 
current density ${\bf J}^{\mu}({\bf r})$ given by 
\begin{equation}
N^{\mu}=\int J^{\mu}e^{i{\bf q}{\cdot}{\bf r}} d^{3}r .
\label{frmfac} 
\end{equation}
Current conservation and gauge invariance can be 
used to eliminate the $z$-components so that only $N_{0}$, $N_{x}$, 
and $N_{y}$ need to be calculated.

By keeping the momentum and energy transfer fixed while varying the  
electron energy $E$ and scattering angle $\theta_e$, or varying the  
azimuthal angle of the outgoing proton and the helicity of the  
incident electron, it is possible to extract from experiment the  
five structure functions as a function of momentum and energy  
transfer.   
However, when the electron wavefunctions are not Dirac  plane waves, 
but rather are distorted by the static Coulomb field of  
the target nucleus, such a simple formulation as given in 
Eq.~(\ref{crosep}) is no longer possible.  In the full Distorted Wave
Born Approximation (DWBA) calculation, it is not possible to express the cross 
section as a sum of bi-linear products of transforms of transition current matrix elements which only depend on the electron kinematics and the outgoing proton's azimuthal angle.  The key point is that the momentum transfer does not enter the analysis in a natural way.   Of course, one could pretend that the plane-wave result is still valid and extract the various "structure functions'.  However, there is no way in the DWBA approach to investigate these terms separately.   

Using partial wave analysis, the Ohio University 
group~\cite{jinphd}--~\cite{jin94}  
has treated the Coulomb distortion arising from the static Coulomb  
field of the nucleus exactly.  
Coupling the distorted waves with the one hard photon exchange approximation 
(Distorted Wave Born Approximation-DWBA) has allowed this analysis 
to be compared to data from a range of nuclei.  
The nuclear model used includes the following  ingredients:  
1) Relativistic Hartree single particle wavefunctions  
for the bound orbitals~\cite{horo81,zhang91}, 
2) Relativistic optical model for the  
continuum proton~\cite{clark90}, and 3) the free relativistic 
current operator for the proton with the standard form factors.  
This simple relativistic ``one-body'' model along with the exact 
treatment of Coulomb distortion is in excellent agreement with all 
the data involving knock-out from surface orbitals of nuclei that have
been analysed.  This includes nuclei as light as   
$^{16}O$ and as heavy as $^{208}Pb$.   
However, for cases where the outgoing particle is in the continuum 
where all multipoles can contribute, the DWBA analysis requires 
extensive computer codes which require more and more time and 
precision as the energy increases.  
Furthermore, as noted above in the exact DWBA analysis, the cross section cannot 
be written as the sum of five terms which are bi-linear products 
of transforms of the transition current matrix elements.  
There two reasons have led researchers to seek an  approximate 
treatment of Coulomb distortion that would permit the extraction 
of ``structure'' functions from experiment under conditions where the effects of 
Coulomb distortion in the different terms can be investigated, and could be easily extended to higher energies such as will be available at the Thomas Jefferson  
National Accelerator Facility.

In a previous paper~\cite{kim1} we investigated a rather extreme 
approximation of Coulomb distortion effects for the inclusive 
reaction $(e,e')$ in the quasielastic region and using some $ad-hoc$ 
assumptions obtained excellent agreement with the exact DWBA 
calculations.  
In this paper, we apply a more exact approximation to the exclusive 
reaction $(e,e'p)$ and investigate its validity.  
One of our advantages as compared to previous investigators is that 
we have the full DWBA calculation to use as a standard in assessing 
the accuracy of our approximation.  
In Section II we will give the approximate electron wavefunctions 
for the incoming and outgoing electron waves and derive the 
approximate M${\o}$ller potential from these wavefunctions.  
In Section III we use the approximate M${\o}$ller potential to calculate the 
differential cross section for $(e,e'p)$ and define the approximate 
Coulomb deformed structure functions.  
In Section IV we compare approximate and DWBA $(e,e'p)$ cross 
sections for so-called parallel and perpendicular kinematics and 
investigate the extraction from the cross sections of the so-called 
fourth ($R_{LT}$) and fifth ($R_{LT'}$) structure functions.  
In Section V we give our conclusions and discuss prospects for 
the future.

\section{Approximate Wavefunctions and the M${\O}$ller Potentials}

Following the work of Lenz and Rosenfelder~\cite{lenz}, we propose 
the following plane-wave-like electron 
wavefunction~\cite{kim1,kimphd} which contains the effects of the 
static Coulomb distortion of the target nucleus in an approximate 
way:
\begin{equation}
\Psi^{({\pm})}({\bf r})={\frac {p'(r)}{p}}
e^{{\pm}i{\delta}({\bf J}^{2})}e^{i \Delta }e^{i{\bf p}^{\prime}(r)
{\cdot}{\bf r}}u_{p}. \label{appwv4}
\end{equation}
The ($\pm$) sign denotes incoming and outgoing boundary conditions for 
electrons of momentum ${\bf p}$, the phase factor 
$\delta({\bf J}^{2})$ is a function of the square of the angular 
momentum operator ${\bf J}$, and the local effective 
momentum ${\bf p}'(r)$
is given in terms of the Coulomb potential of the target nucleus 
by 
\begin{equation}
{\bf p}'(r)=(p-{\frac 1 r} \int_{0}^{r} V(r) dr){\hat {\bf p}} .
\label{lem}
\end{equation}
We refer to this $r$-dependent momentum as the Local Effective 
Momentum Approximation(LEMA).
Some small higher order corrections have been incorporated into the 
${\em ad-hoc}$ term $\Delta = a({\hat p}^{\prime}(r){\cdot}
{\hat r})({\bf J}^{2}+{\frac {1} {4}})$ which involves
 the factor $a$ which is parametrized by 
$a=-{\alpha}Z({\frac {16} {p}})^{2}$ where $Z$ is the charge of 
the target nucleus and the number 16 is given 
in MeV/c and was determined by comparison with the exact result.
Inclusion of this term with LEMA is referred to as LEMA + $\Delta$.
The electron mass has been neglected in comparison to the electron 
momentum, so this approximation if not valid at extreme forward and 
backward electron scattering angles.

Previous workers~\cite{giupac,trani} replaced the $r$-dependent 
momentum $p'(r)$ in Eq.~(\ref{lem}) by the value 
$p'_{EMA}=p-V(0)$.
This approximation is known as the Effective Momentum Approximation 
(EMA). 
Unfortunately, as we have known, the EMA describes the Coulomb 
effects on the wavefunction rather poorly~\cite{kim1,kimphd}. 
Previous workers~\cite{giupac,trani} also approximated the Coulomb 
phases by a constant plus a linear term in the operator 
${\bf J}^{2}$. 
While this approximation works well for low partial waves, it does 
not describe partial waves with angular momenta equal to or greater 
than $pR$ where $R$ is the nuclear radius. 
However, it is these partial waves that dominate inelastic electron 
scattering from the nucleus. We refer to the EMA plus linear fit
to the phases as EMA-$\kappa^2$.
We avoid this problem by fitting the exact partial wave phase shifts 
${\delta}_{\kappa}$, where $\kappa$ is the Dirac quantum number, 
to an expansion in powers of $\kappa^2$.  Retaining terms of second order in $\kappa^2$ for
kappa values up to approximately $3(pR)$, we fit 
the exact phases with the following equation, 
\begin{equation}
{\delta}_{\kappa}=b_{0}+b_{2}{\kappa}^{2}+b_{4}{\kappa}^{4} .
\label{apphs}
\end{equation}
Note that the eigenvalues of ${\bf J}^2$ are $j(j+1)$ which equals ${\kappa}^2 -\frac{1}{4}$.
 
We also investigated the expansion of the phase exponential 
in Eq. (\ref{appwv4}) into a power 
series carried out by previous workers~\cite{giupac,trani} and 
concluded that an accurate description requires 
many terms. 
We chose instead to neglect the electron spin dependence of the 
phases and replace $J^2+\frac{1}{4}$  by the angular momentum squared
$L^2$ in  the exponential phase term for both incoming and outgoing waves.
Further, we replace $L^2$ by its classical value $({\bf r}{\times}{\bf p}^{\prime})^{2}$.
  With these two
approximations, 
the approximate Coulomb wavefunction  is given by
\begin{eqnarray}
\Psi^{({\pm})}({\bf r})&=&{\frac {p'(r)} {p}}
e^{{\pm}ib_{0}}e^{{\pm}ib_{2}({\bf r}
{\times}{\bf p}(r))^{2}}\nonumber\\
&&{\times}e^{{\pm}ib_{4}({\bf r}{\times}{\bf p}(r))^{4}}
e^{ia({\hat p}^{\prime}(r){\cdot}{\hat r})({\bf r}
{\times}{\bf p}(r))^{2}}e^{i{\bf p}^{\prime}(r){\cdot}{\bf r}}u_{p}.
\label{k4led}
\end{eqnarray}
The merits of this approach are that our approximate distorted
Coulomb wavefunctions have an analytic plane-wave-like form and
 the Coulomb distortion modifications do not
depend on the electron spin, but the wavefunctions do have
$({\bf r}{\times}{\bf p'}(r))^{2}$ terms in the exponential which
carry information about the phase shifts.
We will refer to this wavefunction as the approximate analytic Coulomb distorted wavefunction.  
Based on our investigations, it is a good representation of the exact partial wave solutions
for radial coordinates out to about three or four nuclear radii.

Using a technique introduced by Knoll \cite{knoll} which approximates the 
potential in terms of the source current  we  obtain the following approximate M${\o}$ller-type potential corresponding to our approximate analytic Coulomb distorted wavefunction,
\begin{eqnarray}
A^{\mu}({\bf r})&=&{\frac {4{\pi}}
{4p_{i}p_{f} \sin^{2} {\frac {\theta_{e}} {2}}}}
e^{i(b_{0i}+b_{0f})}e^{i[b_{2i}({\bf r}{\times}
{\bf p}_{i}^{\prime}(r))^{2}+b_{4i}({\bf r}{\times}
{\bf p}_{i}^{\prime}(r))^{4}]} \nonumber \\
&{\times}&e^{i[b_{2f}({\bf r}{\times}{\bf p}_{f}^{\prime}(r))^{2}
+b_{4f}({\bf r}{\times}{\bf p}_{f}^{\prime}(r))^{4}]} \nonumber \\
&{\times}&e^{i[a_{i}({\hat r}{\cdot}{\hat p}_{i}(r))({\bf r}
{\times}{\bf p}_{i}^{\prime}(r))^{2}-a_{f}({\hat r}{\cdot}
{\hat p}_{f}(r))({\bf r}{\times}{\bf p}_{f}^{\prime}(r))^{2}]}
\nonumber\\
&{\times}&e^{i{\bf q}^{\prime}(r){\cdot}{\bf r}}{\bar u}_{f}
{\gamma}^{\mu}u_{i}. \label{potk4D}
\end{eqnarray}
The Knoll approach is discussed in detail in  previous work \cite{kim1,kimphd} and is
a good approximation for momentum transfers greater than about 350 MeV/c.  In arriving
at Eq. (\ref{potk4D}), we neglected spatial derivatives of the phase factors in the wavefunction and
the radial derivative of the local effective momenta ${\bf p'}(r)$.  

The comparison of the approximate four potentials with the DWBA four
potential requires a partial wave expansion since that is the only way the
DWBA results can be obtained.
However, the partial wave expansion for these potentials is hardly possible
because of the $({\bf r}{\times}{\bf p}^{\prime})^{2}$ terms in the
exponential.
Traini $et$ $al.$,~\cite{trani} expanded the exponential in a power
series up to the second order so as to use  partial waves, but this
series converges very slowly because the value of the phase,
$b({\bf r}{\times}{\bf p})^{2}$, is greater than one for regions of space 
near the nuclear surface.  Of course, the origin of our approximations was in
a partial wave formalism, so we can go back to that formalism and replace the
exact phases and radial wavefunctions by the approximate phases and wavefunctions to obtain a measure
of the quality of our approximations.  Note, however, that obtaining the 
plane-wave-like form required some additional approximations including assuming that the asymptotic 
momentum transfer ${\bf q}$ and the phase factors in the approximate four potential do not 
explicitly depend on the incoming and outgoing electron spins. 

\begin{figure}[p]
\newbox\figIIa
\setbox\figIIa=\hbox{
\epsfysize=130mm
\epsfxsize=160mm
\epsffile{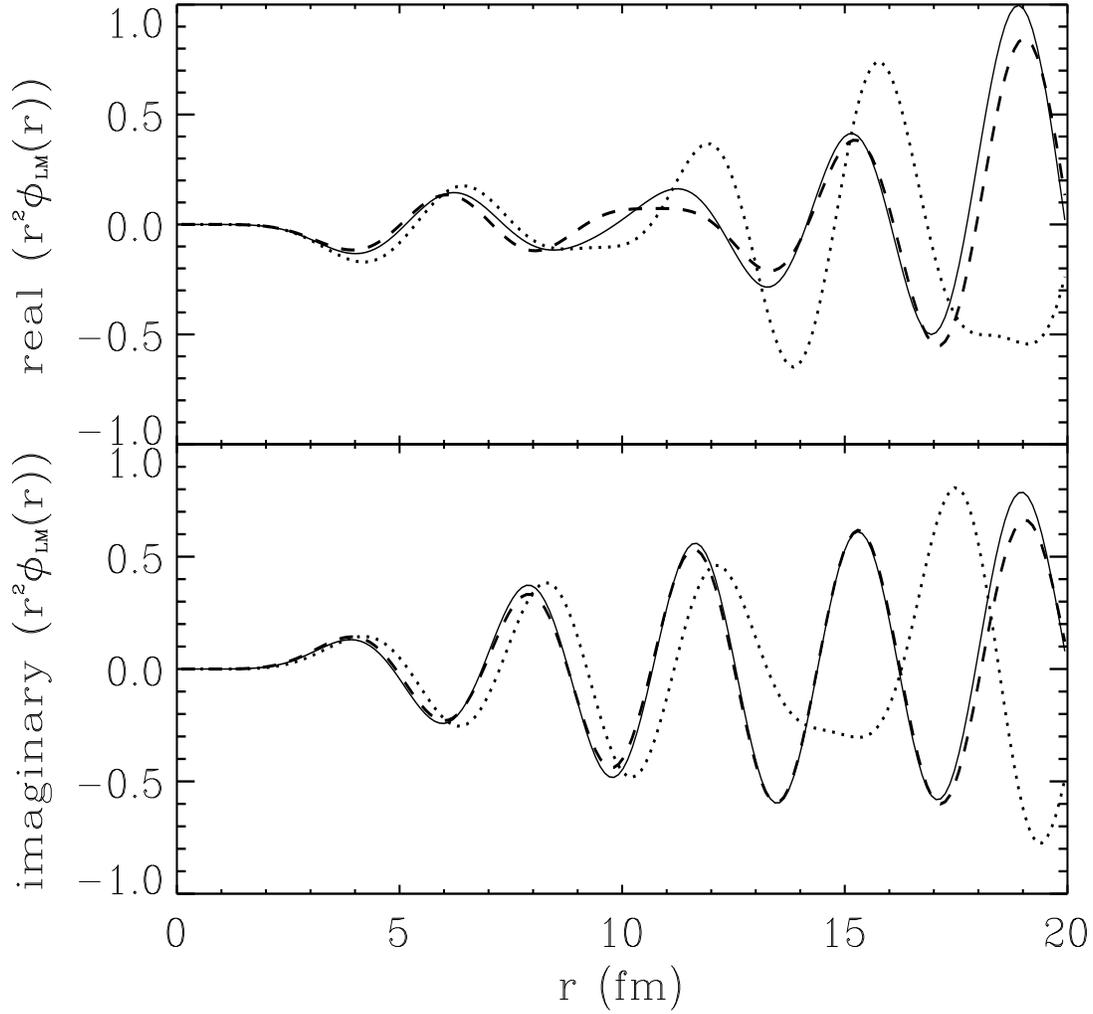}}
\noindent\hspace{0mm}\vspace{20mm}\box\figIIa
\caption{Comparison of the exact scalar potential with approximate scalar potentials for the 
multipole component L=5, M=0. 
The initial energy $E_{i}=400$ MeV and the energy loss 
$\omega=100$ MeV. 
The solid line is calculated with DWBA, the dotted line with $EMA-{\kappa}^{2}$ and 
the dashed line is our approximate potential.}
\label{scalar5}
\end{figure}

\begin{figure}[p]
\newbox\figIIb
\setbox\figIIb=\hbox{
\epsfysize=130mm
\epsfxsize=160mm
\epsffile{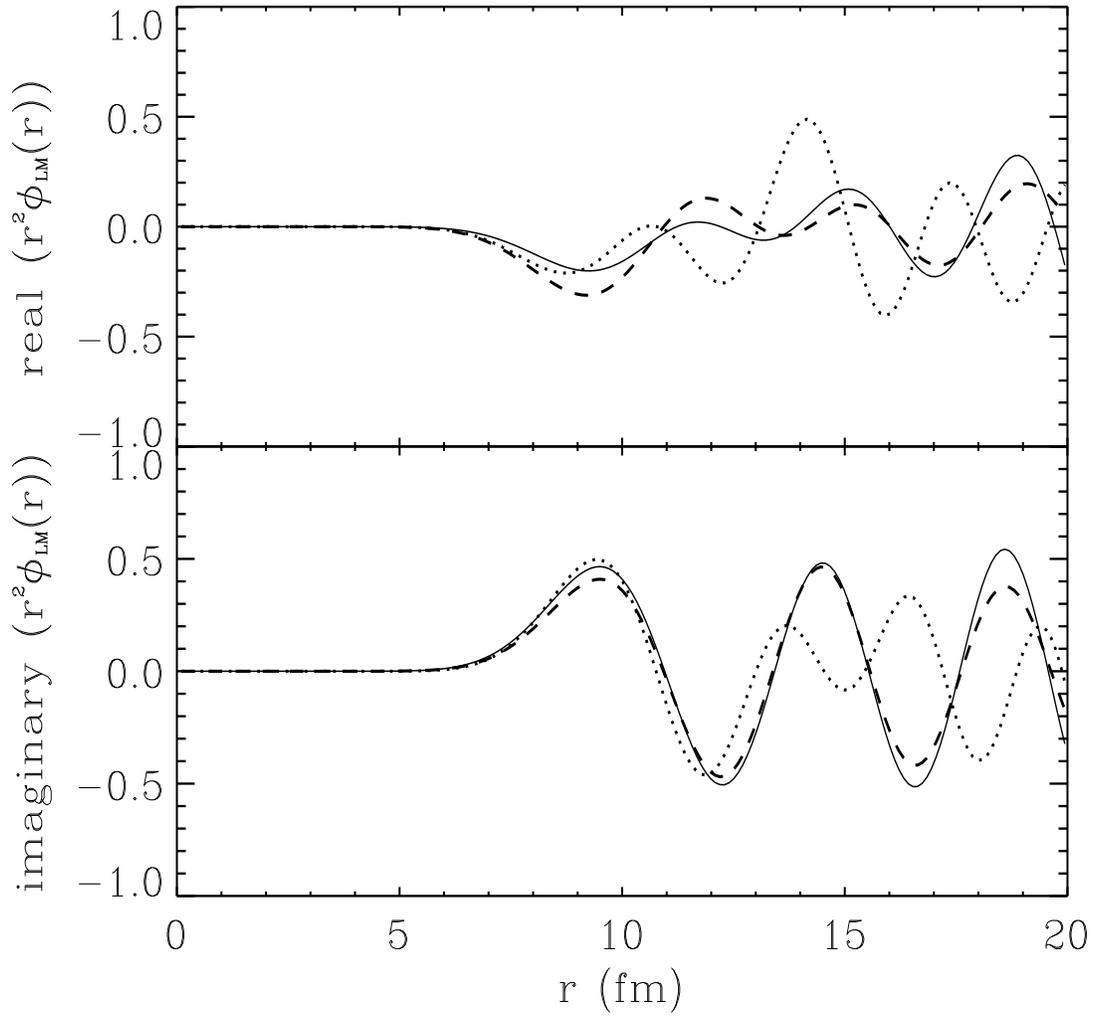}}
\noindent\hspace{0mm}\vspace{20mm}\box\figIIb
\caption{The same as Fig.~(\ref{scalar5}) with multipole $L=15$ 
and $M=0$.}
\label{scalar15}
\end{figure}

With this caveat, we can compare various approximate potentials that
have been widely used ~\cite{giupac,trani}
and our approximate potential for selected multipoles. 
Figs.~(\ref{scalar5}) and ~(\ref{scalar15}) show the comparison of 
two scalar potentials with the DWBA scalar potential by using
the partial waves for different multipole $L$ values. The
initial spin and the final spin of the electron are
$s_{i}=s_{f}=1/2$.
The calculations have been done using a Fermi charge distribution of
radius $R=6.65$ fm, with total charge $Z=82$ and the angular
momentum $L=5$ and $15$.
The initial electron energy is $E_{i}=400$ MeV, the final
electron energy is $E_{f}=300$ MeV, and the momentum transfer is
$q=350.5$ MeV/c.
The solid line is the result for the DWBA scalar potential, the
dotted line is for the $EMA-{\kappa}^{2}$ potential, and the dashed
line is for our approximate potential of Eq. (\ref{potk4D}).
The approximation for $EMA-{\kappa}^{2}$ has a different magnitude
and is also out of phase, but that for our approximate potential almost
has the same magnitude and is in phase.
The discrepancy at large radii is due to lack of complete convergence
of the $\kappa$ series in the DWBA calculation and does not play an important role in
calculating the cross section since the bound nucleon wavefunction
drops very rapidly with radial distance.  We conclude that our approximate potential
is in quite good agreement with the exact potential calculated with partial waves (DWBA)
for radial coordinates less than three to four nuclear radii, and hence
 is good enough to replace
the full DWBA calculation.
We have confirmed with the aid of a simple model\cite{kimphd} that our approximate
potential reproduces the cross section calculated with the full partial wave result from DWBA
quite well.  We will show more realistic comparisons in the following sections that do
not utilize a multipole decomposition and therefore is a more direct test of the
approximate Coulomb distorted potential given in Eq.~(\ref{potk4D}).

\section{Approximate Coulomb Distorted Cross Section}

Using the approximate M${\o}$ller potential given in Eq.~(\ref{potk4D}) it is straightforward
to derive the cross sections for $(e,e'p)$ reactions from nuclei since apart from
the modified spatial dependence the approximate potential has the same
Dirac structure as the plane wave M${\o}$ller potential.  The result is
\begin{eqnarray}
{\frac {d^{3}{\sigma}} {dE_{f}d{\Omega}_{f}d{\Omega}_{P}}}
&=&{\frac {PE_{P}} {(2{\pi})^{3}}}{\sigma}_{M}[v_L R^{\prime}_{L}+
 v_T R^{\prime}_{T}+
{\cos {2{\phi}_{P}}} v_{TT} R^{\prime}_{TT} \nonumber \\
&+&{\cos {{\phi}_{P}}} v_{LT} R^{\prime}_{LT}
+h {\sin {{\phi}_{P}}} v_{LT^{\prime}}R^{\prime}_{LT^{\prime}}]. \label{apcross}
\end{eqnarray}
The electron structure functions are unchanged from Eq.~(\ref{elecst}), but the
nuclear structure functions, designated with a prime superscript, contain Coulomb
distortion effects.  They are defined as in Eqs. (\ref{structure})
and (\ref{nclten}), except that the ``Fourier'' transforms
given in Eq.~(\ref{frmfac}) become
\begin{eqnarray}
N^{\prime}_{0}&=&\int ({\frac {q'_{\mu}(r)} {q_{\mu}}})^{2} 
({\frac {q} {q'(r)}})^{2} e^{i (\delta_i + \delta_f + \Delta_i - \Delta_f)}
 J_{0}e^{i{\bf q}^{\prime}(r){\cdot}{\bf r}} d^{3}r  \nonumber \\
N^{\prime}_{x,y}&=&\int e^{i (\delta_i + \delta_f + \Delta_i - \Delta_f)}
 J_{x,y}e^{i{\bf q}^{\prime}(r){\cdot}{\bf r}} d^{3}r,
\label{dfrmfac} 
\end{eqnarray}
where the phase shift $\delta$ and the ad-hoc additional phase 
$\Delta$ are functions of $\bf r$ given by
\begin{eqnarray}
\delta &=& b_{0}+b_{2}({\bf r}{\times}
{\bf p}^{\prime}(r))^{2}+b_{4}({\bf r}{\times}
{\bf p}^{\prime}(r))^{4} \nonumber \\
\Delta &=& a({\hat r}{\cdot}{\hat p}(r))({\bf r}
{\times}{\bf p}^{\prime}(r))^{2}. \label{phases}
\end{eqnarray}

The result in Eq. (\ref{apcross}) is our primary finding. Our approximate treatment of 
Coulomb distortion leads to a ``plane-wave-like'' form for the cross section and thereby opens up the possibility of investigating the various "structure functions" independently.  
Of course, these more generalized ``structure functions'' do contain some dependence 
on the electron kinematics, but with the use of theoretical nuclear models, 
this modification of the structure functions can be investigated.

 The charge transform in Eq. (\ref{apcross}) 
differs from the transverse current transforms since the
continuity equation was used to eliminate the z-component of the current.
Note that unlike the case for electron plane waves, the various current transforms
are not azimuthally symmetric about the momentum transfer direction 
$\bf q$, and therefore contain dependencies on the outgoing nucleon
azimuthal angle $\phi_P$ over and beyond the explicit dependencies shown in
Eq.~(\ref{apcross}).  However, some symmetry remains since both $({\bf r}{\times}
{\bf p_i}^{\prime}(r))^{2}$ and $({\bf r}{\times}
{\bf p_f}^{\prime}(r))^{2}$ are invariant under the transformation $\phi \rightarrow -\phi$
which results in the nuclear structure being invariant under 
$\phi_P \rightarrow -\phi_P$.  The consequences of this additional dependence on $\phi_P$
will be discussed further in the next section.

\section{Results}
\subsection{Cross section in parallel and perpendicular kinematics}

In our analysis we are looking at one particular shell, and trying to
find the reduced cross section ${\rho}_{m}$, which for proton
waves in the final state is related to the probability that a bound
proton from a given shell with the missing momentum ${\bf p}_{m}$ can
be knocked out of the nucleus with asymptotic momentum ${\bf P}$.
The reduced cross section as a function of $p_{m}$ is commonly  defined by
\begin{equation}
{\rho}_{m}(p_{m})={\frac 1 {PE_{P}{\sigma}_{eP}}}{\frac
{d^{3}{\sigma}} {dE_{f}d{\Omega}_{f}d{\Omega}_{P}}}, \label{rdcro}
\end{equation}
where the missing momentum can be determined by the kinematics ${\bf
p}_{m}={\bf P}-{\bf q}$.
The off-shell electron-proton cross section ${\sigma}_{eP}$ is not
uniquely defined, but in all cases we use the form 
${\sigma}^{cc1}_{eP}$ given by deForest~\cite{defor83}.

There are two kinematical situations commonly
used in $(e,e'p)$ experiments.
They are parallel kinematics where the outgoing proton momentum
${\bf P}$ is along the momentum transfer ${\bf q}$ and 
perpendicular kinematics where the magnitude of ${\bf P}$ is
fixed, but the detected proton makes an angle ${\theta}_{Pq}$ with
respect to ${\bf q}$.  In perpendicular kinematics, the magnitude of ${\bf P}$
is usually equal to the magnitude of ${\bf q}$.
All calculations will be carried out in the laboratory
frame (target fixed frame).
In the parallel case, the three interference terms in
Eq.~(\ref{rdcro}) disappear, while in the perpendicular case, all
terms remain except the fifth structure function which sums to zero
for unpolarized incoming electrons.
Our approximate calculations for the $(e,e'p)$ reaction include the
approximate phase factors and the correction term $\Delta$ by keeping
the exponential form and the transition matrix element is evaluated
by three dimensional integration since a multipole expansion is no longer
practical.
We compare our results to the full DWBA
results~\cite{jinphd,jineep92} and the experimental data from
NIKHEF~\cite{nikh1,nikh5} in Amsterdam.
The electron incoming energy is given by $E_{i}=412$ MeV and the ejected
proton kinetic energy $T_{P}=100$ MeV.
All calculations include the proton final state interaction by using
a relativistic optical potential obtained from fitting to elastic
proton scattering data~\cite{clark90}.

\begin{figure}[p]
\newbox\figa
\setbox\figa=\hbox{
\epsfysize=130mm
\epsfxsize=160mm
\epsffile{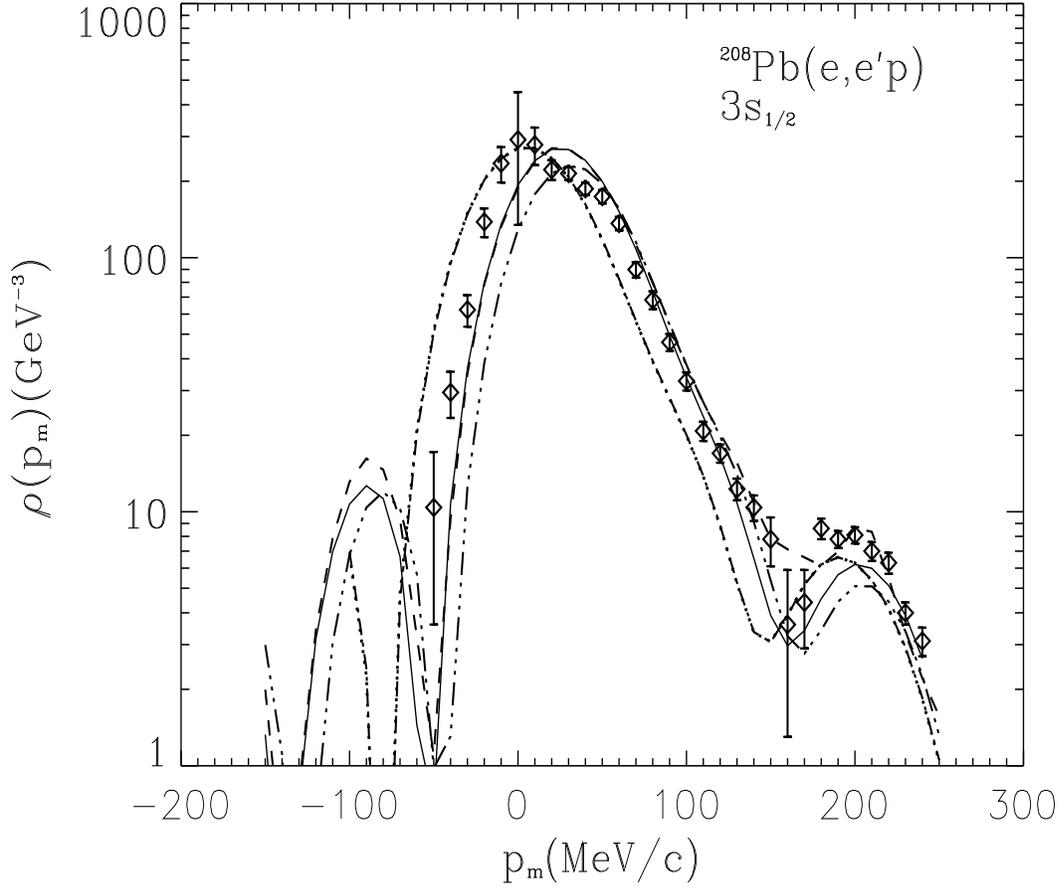}}
\noindent\hspace{0mm}\vspace{20mm}\box\figa
\caption{Reduced cross sections for $^{208}Pb(e,e'p)$ from the
$3s_{1/2}$ shell with parallel kinematics.
The kinematics are $E_{i}=412$ MeV, and proton kinetic energy
$T_{P}=100$ MeV. The dotted line is the PWBA result and the dash-dotted
line (which falls on top of it) is the approximate DWBA  result with $Z=1$. The
dash-three-dotted line is the $EMA-{\kappa}^{2}$ result,
the solid line is the approximate DWBA result,  the dashed line
is the full DWBA result, and the diamonds are data from NIKHEF.
 The same spectroscopic factor of 71\% is used in all curves.}
\label{pars412}
\end{figure}

\begin{figure}[p]
\newbox\figb
\setbox\figb=\hbox{
\epsfysize=130mm
\epsfxsize=160mm
\epsffile{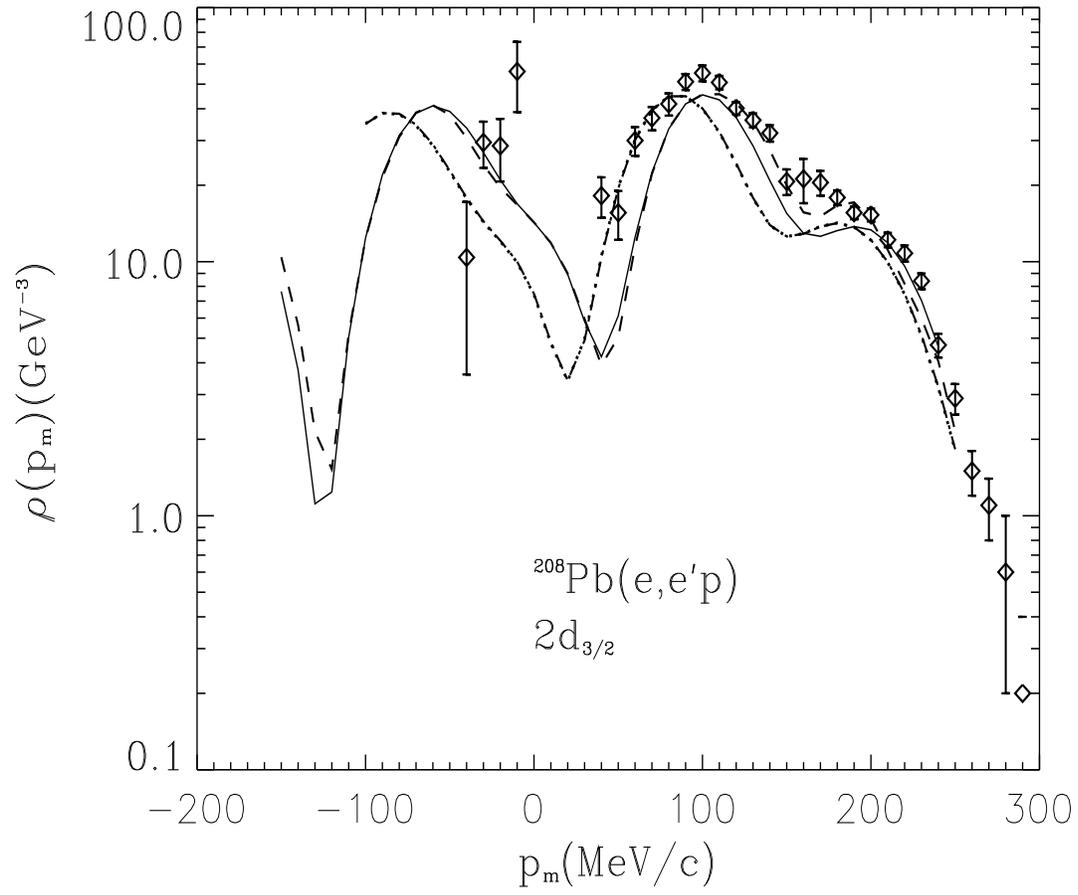}}
\noindent\hspace{0mm}\vspace{20mm}\box\figb
\caption{The same as Fig.~(\ref{pars412}) except for the $2d_{3/2}$ 
shell and the $EMA-{\kappa}^{2}$ result is not shown.}
\label{pard412}
\end{figure}

In Figs.~(\ref{pars412}) and~(\ref{pard412}), we show two results
corresponding to knocking out a proton from a $3s_{1/2}$ shell and a
$2d_{3/2}$ shell in $^{208}Pb$ for the case of parallel kinematics.
For this kinematics, the proton momentum ${\bf P}$ is parallel to the
asymptotic momentum transfer ${\bf q}$ which defines the
${\hat {\bf z}}$ axis and the missing momentum ${\bf p}_{m}$ is also along
the ${\bf q}$ direction.
The dotted line is the PWBA result obtained by using a multipole
expansion and doing the one dimensional integration over $r$ in the
normal way, while the dash-dotted line uses the approximate
potential with Z=1 evaluated by three dimensional numerical integration.  They are
in excellent agreement (to better than $1\%$) as they should be since
the approximate calculations approaches the plane wave result as $Z{\rightarrow}0$.
The solid line is our approximate Coulomb distorted result obtained by numerical integration while the dashed line is the full DWBA results obtained by using partial waves and multipole analysis.  The dash-three-dotted line is the
 $EMA-{\kappa}^{2}$ result also
obtained by using three dimensional integration.    
The diamonds are data from NIKHEF.  Note that the primary effect of Coulomb distortion 
is to shift the reduced cross section by about 20 MeV/c in missing
momentum to the right as compared to the plane wave results.
The approximate DWBA results reproduce the full DWBA results well
around the first peak where the difference is less than $2\%$,
but deviate somewhat for large missing momentum where the reduced cross section is smaller.
The approximate DWBA result breaks down rapidly beyond
missing momentum $p_{m}=100$ MeV/c on the right side, but
since ${\bf q}={\bf P}-{\bf p}_{m}$,  positive
$p_{m}$ corresponds to small $q$ and we expect our approximation to
become worse for $q$ less than about 350 MeV/c as discussed in
Section 2.

\begin{figure}[p]
\newbox\figc
\setbox\figc=\hbox{
\epsfysize=130mm
\epsfxsize=160mm
\epsffile{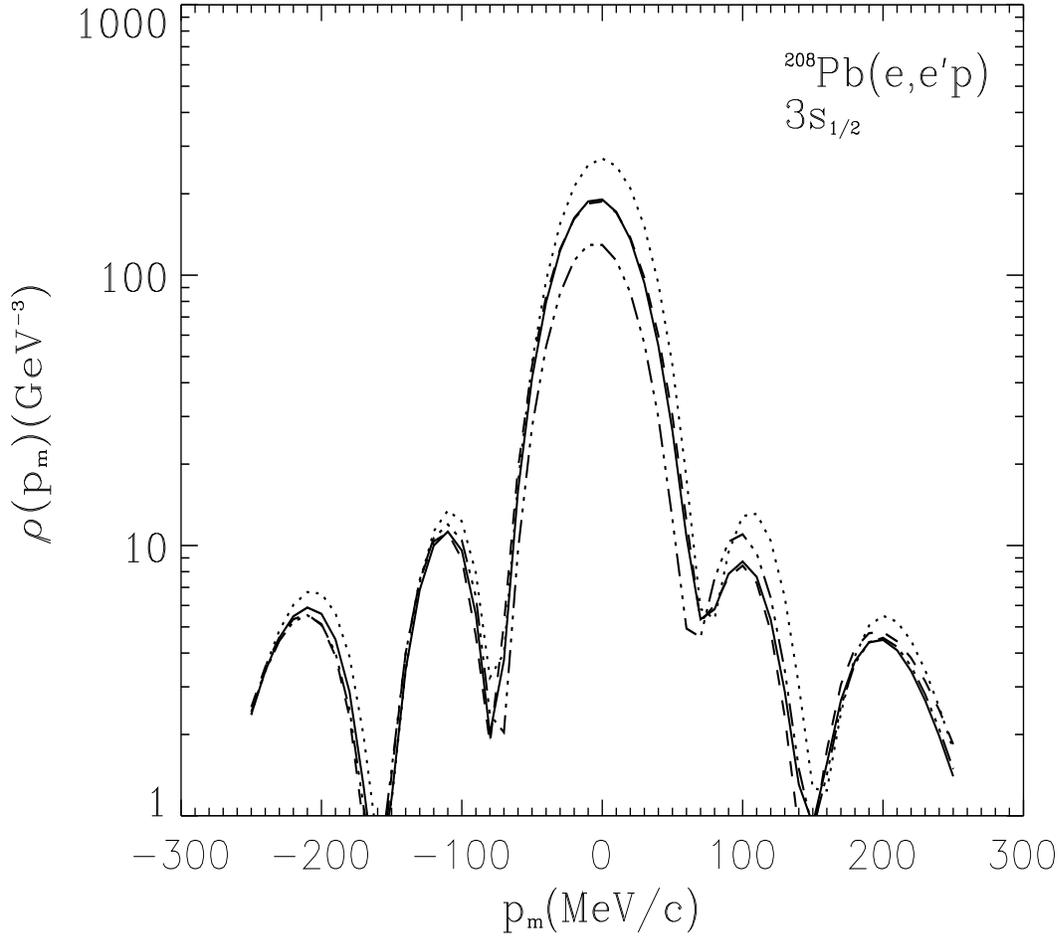}}
\noindent\hspace{0mm}\vspace{20mm}\box\figc
\caption{Reduced cross sections for $^{208}Pb(e,e'p)$ from the
$3s_{1/2}$ shell with perpendicular kinematics.
The kinematics are $E_{i}=412$ MeV, and proton kinetic
energy $T_{P}=100$ MeV. The dotted line is the PWBA result, the
dash-three-dotted line is the $EMA-{\kappa}^{2}$ result,
the solid line is our approximate result,
and the dashed line is the full DWBA result.}
\label{pers412}
\end{figure}

\begin{figure}[p]
\newbox\figd
\setbox\figd=\hbox{
\epsfysize=130mm
\epsfxsize=160mm
\epsffile{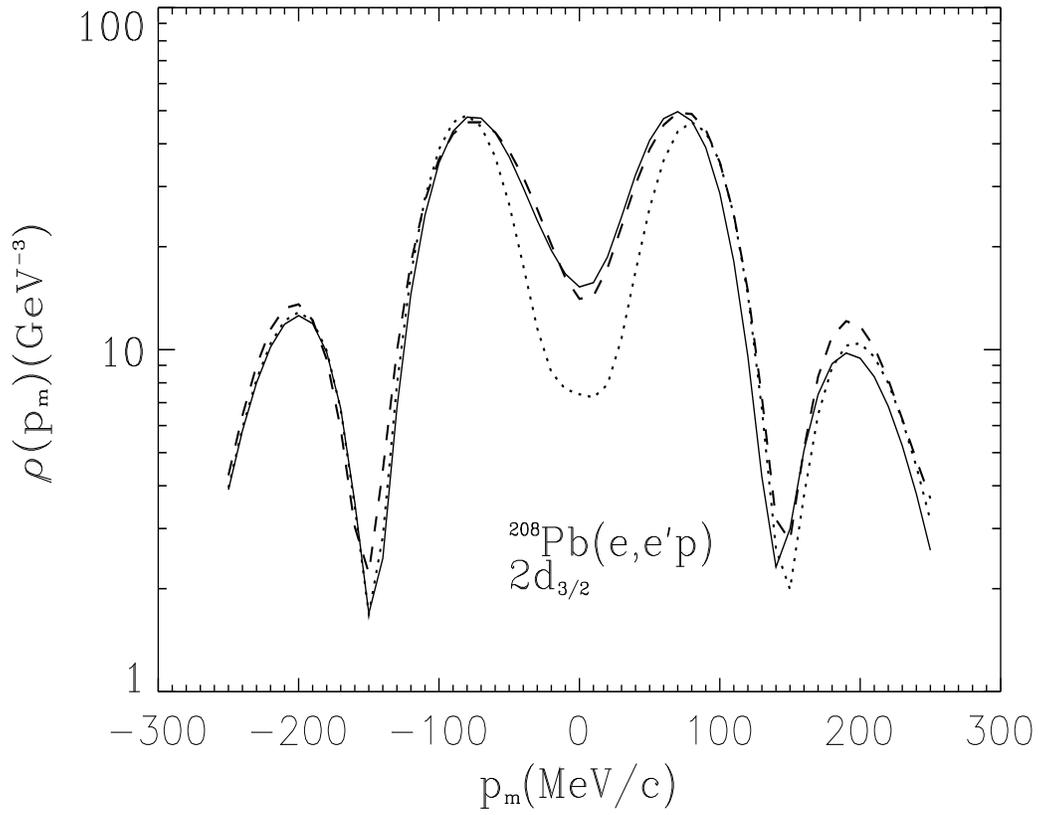}}
\noindent\hspace{0mm}\vspace{20mm}\box\figd
\caption{The same as Fig.~(\ref{pers412}) except the $2d_{3/2}$ 
shell.}
\label{perd412}
\end{figure}

The $EMA-{\kappa}^{2}$ result is lower by about $30\%$ around the first peak
than the full DWBA result.
Note further that the electron distortion affects the positive
missing momentum $p_{m}$ and the negative $p_{m}$ differently.
The negative $p_{m}$ region shows a large Coulomb
distortion effect.
Figs.~(\ref{pers412}) and ~(\ref{perd412}) show the reduced cross
sections from $3s_{1/2}$ and $2d_{3/2}$ for $^{208}Pb$ for
perpendicular kinematics.
We choose ${\bf P}={\bf q}$ which marks the top of the
quasielastic peak for s-states.  The electron angle corresponding to 
this momentum transfer is
${\theta}_{e}=74^{0}$ for $E_{i}=412$ Mev and the ejected proton
energy $T_{P}=100$ MeV.
The dotted line is the PWBA result, the solid line is our approximate
 result, the dash-three-dotted line is the $EMA-{\kappa}^{2}$
result, and the dashed line is the full DWBA result.
Since the momentum transfer is large,
$q=444$ MeV/c, the approximate  results are in much better
agreement with the full DWBA results than those of the parallel 
kinematics case discussed above.
The difference is less than $2\%$ around the first peak as in
the parallel case, and for both side regions the deviation is around $5\%$. The
positions of the maxima and minima are in the right places.
The discrepancy between the DWBA result and the $EMA-{\kappa}^{2}$
result is greater than $30\%$ around the first peak.
These results confirm previous observations that Coulomb distortion has smaller effects
in perpendicular kinematics than in parallel kinematics.

From these calculations of two different kinematic cases for
the $(e,e'p)$ reaction, the approximate DWBA results reproduce the
full DWBA and the experimental results quite well, especially around
the first peak.
The effect of the Coulomb distortion on the cross section for
a knocked-out proton from the
$3s_{1/2}$ shell for $^{208}Pb$ is almost $30\%$ but that for
knocked-out from the $2d_{3/2}$ shell is only $10\%$ as compared
to the PWBA calculation.

\begin{figure}[p]
\newbox\figg
\setbox\figg=\hbox{
\epsfysize=130mm
\epsfxsize=160mm
\epsffile{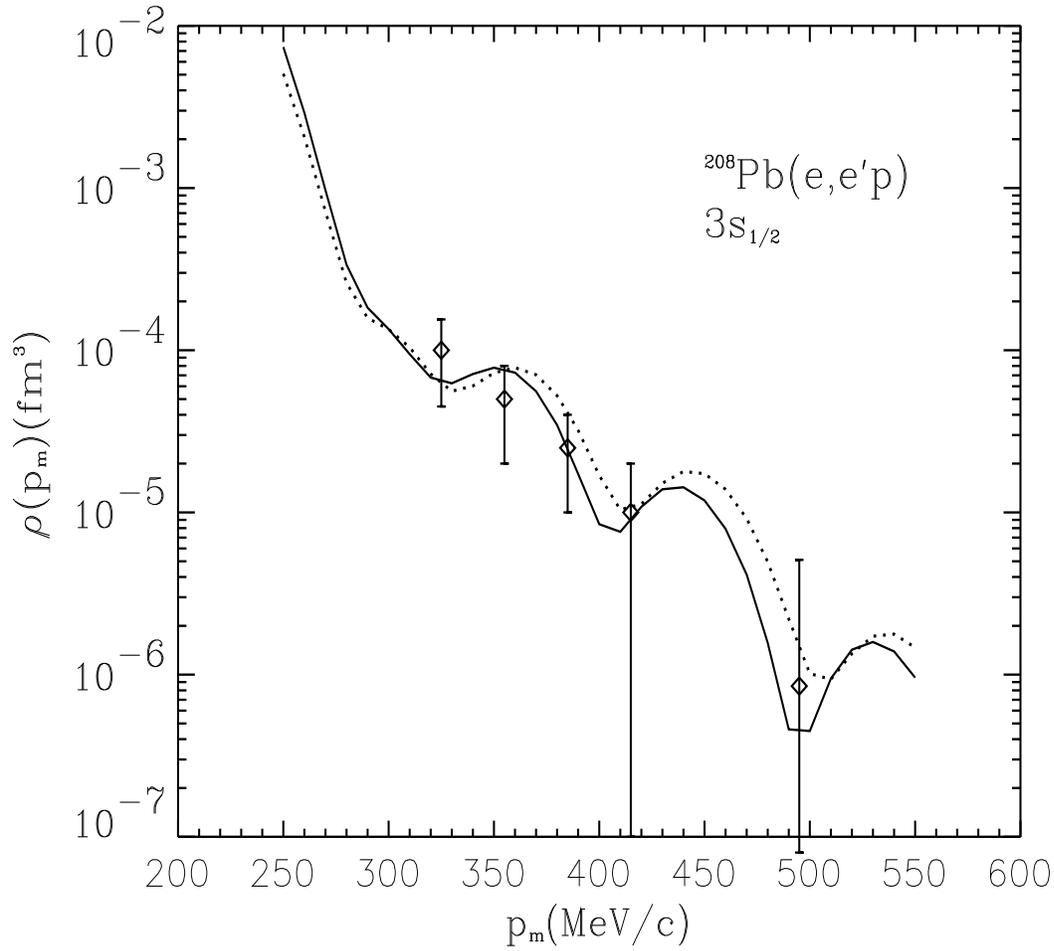}}
\noindent\hspace{0mm}\vspace{20mm}\box\figg
\caption{Reduced cross sections for $^{208}Pb(e,e'p)$ from the
$3s_{1/2}$ shell for high missing momentum.
The kinematics are $E_{i}=487$ MeV, momentum transfer $q=221$ MeV/c,
energy transfer $\omega=110$ MeV, and proton kinetic energy
$T_{P}=100$ MeV.
The solid line is the approximate DWBA result, the dotted
line is the PWBA result, and the diamonds are data from NIKHEF.  A 
previously determined spectroscopic factor of 71\% was used for 
both curves.}
\label{high}
\end{figure}

In the past, $(e,e'p)$ experiments in parallel kinematics have been
measured for the reduced cross section in the missing momentum
range $-50{\leq}p_{m}{\leq}300$ MeV/c at NIKHEF.
Recently, the range of the missing momentum has been extended to
$300{\leq}p_{m}{\leq}500$ MeV/c by $(e,e'p)$ measurements for
perpendicular kinematics~\cite{bob94}.
The new reduced cross section was measured at momentum transfer
$q=221$ MeV/c, energy transfer $\omega=110$ MeV, the kinetic energy
of the detected proton $T_{p}=100$ MeV, and incident electron energy
$E_{i}=487$ MeV as shown in Fig.~(\ref{high}).
The dotted line is the PWBA result, the solid line is our approximate
DWBA result, and the diamonds are the experimental data from
NIKHEF~\cite{bob94}.
Our result reproduces the measured reduced cross section very well
although the momentum transfer $q$ is small.  A similar conclusion
has beed drawn~\cite{madrid} by the Madrid group using their DWBA calculation
to analyse this same data.
The interesting physics point is that our ``single-particle'' relativistic
model reproduces the experimental data at large missing momentum
quite well and should be compared to an analysis of this same reaction with a 
non-relativistic
approach which uses a non-relavistic current operator and
 finds it necessary to introduce many different two-body currents to
even come close to the data \cite{Belgium}.
It should be noted that our calculation only contains one free
parameter, the spectroscopic factor which is an overall scale factor of $0.71$
which had already been determined~\cite{jinphd}--~\cite{jin94} by the low missing momentum data.

\subsection{Interference Structure Functions}

In previous work~\cite{jin94}, the effect of Coulomb
distortion on the magnitude of the fourth structure function was more
than $15\%$ in $^{16}O$, and more than a factor of 2 in $^{208}Pb$.
The magnitude effect on the fifth structure function depends on the
out-of-plane angle of the knocked out proton used in the extraction
and was more than $15\%$ for a small angle (e.g.
$10^{0}$) data in $^{16}O$.
Since the structure functions appear in the cross section with
different electron kinematic factors, one can study them
independently, but we will show below that
the PWBA formalism is no longer valid in the
presence of the static Coulomb field of the nucleus.
Even though the separation for the full DWBA calculation with a
partial wave expansion is not valid in the presence of the Coulomb
distortion, it is possible to calculate the fourth and the fifth
structure functions which embody left-right and up-down asymmetries
of the cross section measured with respect to the momentum transfer
direction.
We call a quantity so determined the $apparent$ structure function, and
note that it
would correspond to a structure function extracted from experiment.  In our model,
we can also directly calculate the "structure functions" as given by Eq.~(\ref{structure}) when
the distorted "Fourier" transforms of Eq.~(\ref{dfrmfac}) are used.
One question is to what extent these two results agree with each other.

From Eq.~(\ref{crosep}) one can see that the fourth structure
function $R_{LT}$ could be obtained experimentally by subtracting
the cross sections with ${\phi}_{P}=0$ and ${\phi}_{P}={\pi}$ while
keeping other electron and proton kinematic variables fixed.
The fourth apparent structure function determined by the left-right
asymmetry with respect to the momentum transfer direction is given by
\begin{equation}
R^{a}_{LT}={\frac {{\sigma}^{L}-{\sigma}^{R}} {2Kv_{LT}}} \label{jinlt}
\end{equation}
where $L$ (left) indicates  ${\phi}_{P}=0$ and $R$ (right)
indicates ${\phi}_{P}={\pi}$.  The constant $K={\frac{PE_P}{{2\pi}^3}\sigma_M}$
and the electron structure functions $v$ are defined in Eq.~(\ref{elecst}).
The superscript $a$ means the apparent structure function including
the electron Coulomb distortion, and corresponds to what one would extract 
from experiment.
If the incoming electron beam is polarized ($h=1$), one can extract
the apparent fifth structure function by the up-down asymmetry of the
cross section given by
\begin{equation}
R^{a}_{LT'}={\frac {{\sigma}^{U}-{\sigma}^{D}} {2Kv_{LT'}{\sin
{\phi}_{P}}}} \label{jinltp}
\end{equation}
where $U$ (up) means  $0<{\phi}_{P}<{\pi}$ (above the plane)
and $D$ (down) means  $-{\phi}_{P}$ (below the plane).
 
\begin{figure}[p]
\newbox\figh
\setbox\figh=\hbox{
\epsfysize=130mm
\epsfxsize=160mm
\epsffile{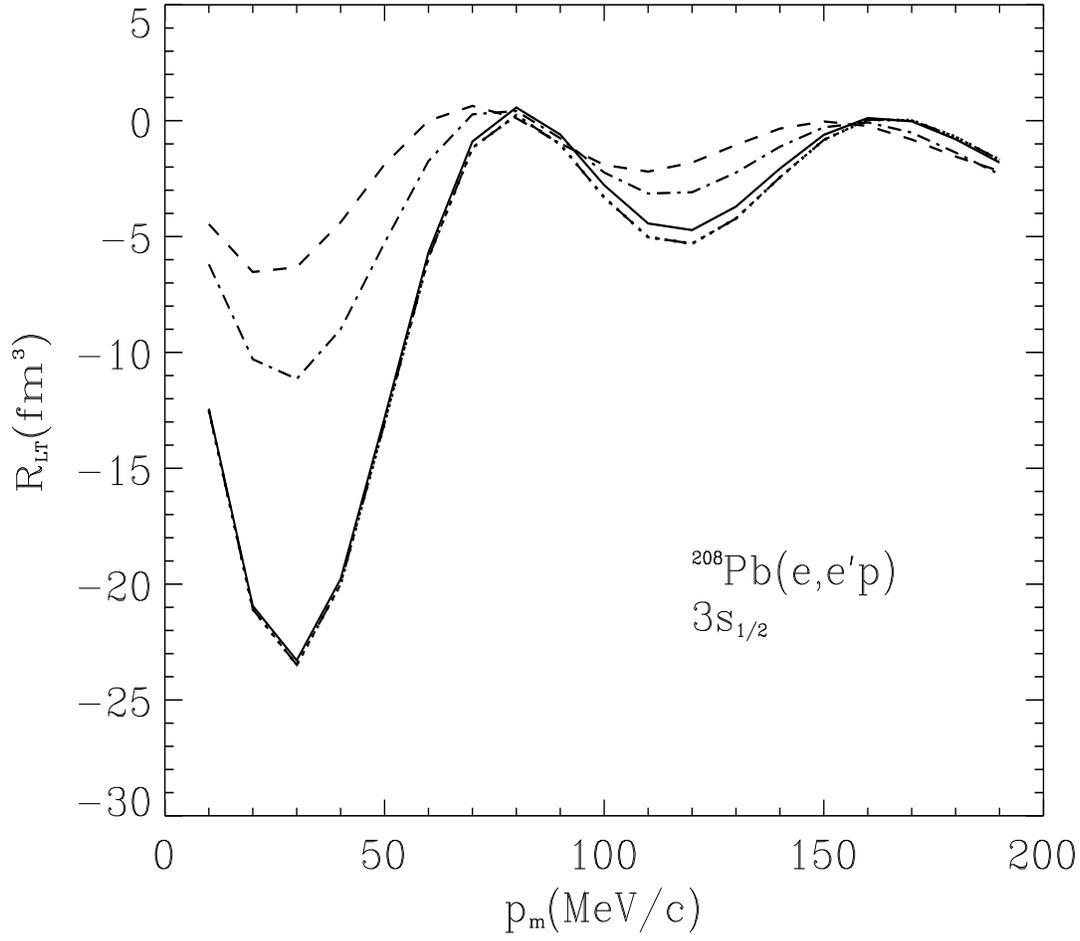}}
\noindent\hspace{0mm}\vspace{20mm}\box\figh
\caption{The fourth structure function for $^{208}Pb(e,e'p)$ from the
$3s_{1/2}$ shell as a function of missing momentum. The kinematics
are $E_{i}=500$ MeV, and proton kinetic energy $T_{P}=100$ MeV. The
dash-dotted line is the apparent structure function and dotted line
is the directly calculated structure function for the PWBA (which fall
on top of each other).  The solid
line is the apparent and the dashed line is the directly calculated
structure function for the approximate DWBA result, while the
dash-three-dotted line is the apparent structure function for the
full DWBA result.}
\label{fours500}
\end{figure}

We extract the fourth structure functions for the $3s_{1/2}$ orbit
of $^{208}Pb$ with incident electron energy $E_{i}=500$ MeV
as shown Fig.~(\ref{fours500}).
The dotted line and the dashed line are the fourth structure
functions calculated directly using  Eq.~(\ref{dfrmfac}) for the PWBA
and the approximate DWBA result, while the dash-dotted line,
the solid line, and the dash-three-dotted line are the apparent
fourth structure functions from Eq.~(\ref{jinlt}) for the PWBA, the
approximate DWBA and the full DWBA calculation, respectively.
We compared our approximate DWBA calculation of the
cross section  to the full DWBA and found the difference to be
less than $2\%$ around the first peak, and  around $5\%$ at the
second peak.
When Coulomb distortion
is included the directly calculated fourth structure function differs
from the apparent fourth structure function by a factor of
3 at the peaks.
Of course for the PWBA calculation the apparent fourth structure
function agrees exactly with the directly calculated fourth structure
function.
Clearly, the standard separation formalism is no longer valid in the
presence of the electron Coulomb distortion for the fourth structure function
$R_{LT}$.  Furthermore, while the
effect of the electron Coulomb distortion is on the order of
$30\%$ for the cross section, it changes  the apparent fourth structure
function by more than a factor of 2.

\begin{figure}[p]
\newbox\figi
\setbox\figi=\hbox{
\epsfysize=130mm
\epsfxsize=160mm
\epsffile{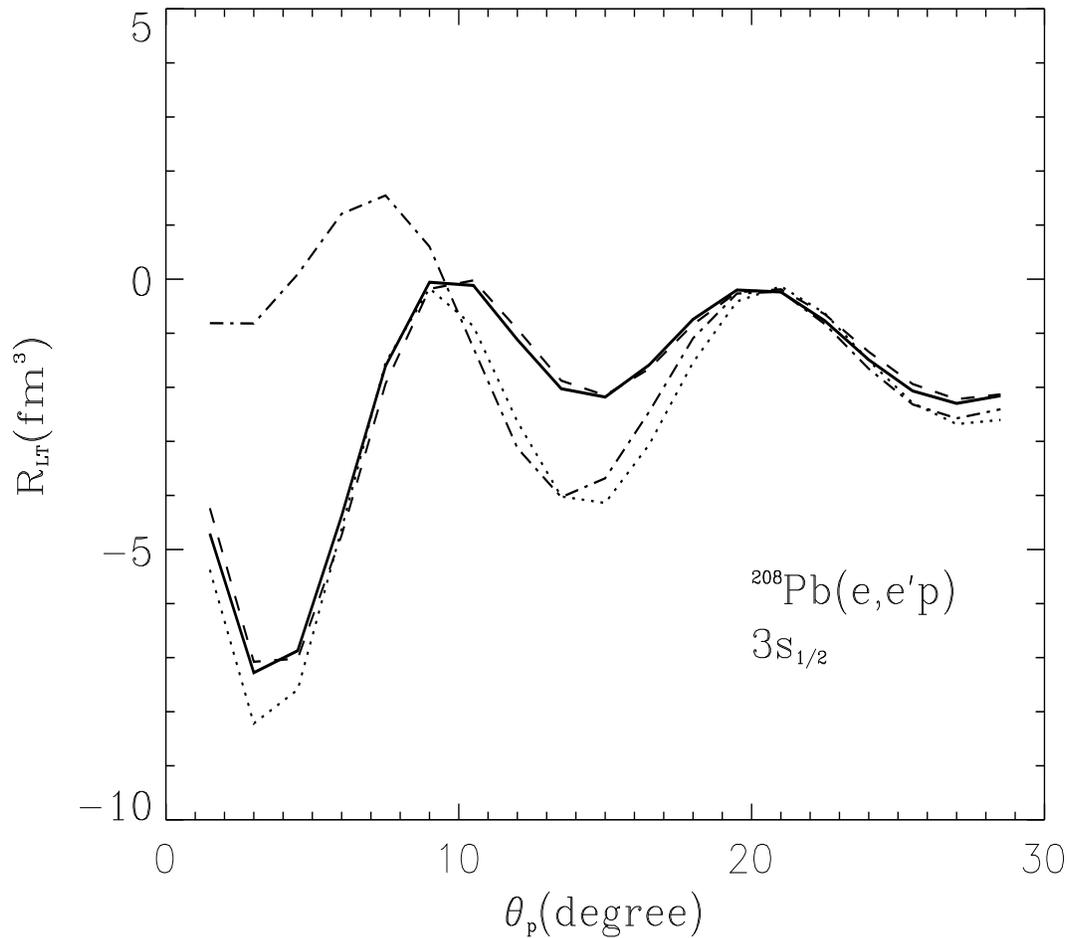}}
\noindent\hspace{0mm}\vspace{20mm}\box\figi
\caption{The fourth structure function for $^{208}Pb(e,e'p)$ for
the $3s_{1/2}$ orbit as a function of the polar angle of the ejected
proton. The solid line and the dotted line are the direct
 and the apparent structure function with the
${\hat {\bf z}}$ axis along the momentum transfer ${\bf q}'(R)$,
and the dashed line and the dash-dotted line are the direct and the apparent
structure function for the case of the ${\hat {\bf z}}$ axis
along the EMA momentum ${\bf q}'(0)$.}
\label{mfour500}
\end{figure}

In Fig.~(\ref{fours500}) the discrepancy between the directly
calculated fourth structure function $R'_{LT}$ and the apparent fourth structure
function for the approximate DWBA suggests that the structure functions
depend on the azimuthal angle of the ejected proton as expected. 
In order to reduce this dependence, we investigated changing
the definition of  the ${\hat {\bf z}}$ axis, 
normally defined by asymptotic momentum transfer ${\bf q}$, in order to
bring the apparent and direct structure functions into closer agreement.
 We considered two choices, one where ${\hat {\bf z}}$ is taken to lie along
${\bf q}'(R)={\bf p}'_{i}(R)-{\bf p}'_{f}(R)$ where
$p'(R)=p-V(R)$ and $V(R)$ is the value of the Coulomb potential at the 
nuclear surface, and the second along
${\bf q}'(0)={\bf p}'_{i}(0)-{\bf p}'_{f}(0)$ where $p'(0)=p-V(0)$  and $V(0)$ is the
Coulomb potential at the origin.  The second case corresponds to the EMA approximation.
We carried out our approximate DWBA calculations for both choices and show the
results 
for the $3s_{1/2}$ orbit of $^{208}Pb$ in
Fig.~(\ref{mfour500}).
The incident electron energy $E_{i}=500$ MeV, the proton kinetic
energy  $T_{p}=100$ MeV, and the outgoing proton momentum is equal
to the momentum transfer q.
The solid line is the directly calculated $R'_{LT}$
and the dotted line is the apparent fourth structure function obtained
when ${\hat {\bf z}}$ is along
${\bf q}'(R)$.   The dashed line is the directly calculated $R'_{LT}$ and the dash-dotted line is the apparent fourth
structure function obtained by using ${\bf q}'(0)$ to define the ${\hat {\bf z}}$ axis.

When choosing the ${\hat {\bf z}}$ axis along ${\bf q}'(0)$, the apparent structure function is out of phase
with the direct structure function and the magnitude is
suppressed at the first peak, but by using the ${\hat {\bf z}}$ axis along ${\bf q}'(R)$ the structure
functions are in phase and the magnitudes are quite close.
Thus, changing the ${\hat {\bf z}}$ axis to be along the direction of 
${\bf q}'(R)$ permits the extraction of a fourth
structure function $R^{\prime}_{LT}$.  Furthermore, choosing this different ${\hat {\bf z}}$ axis
largely removes the Coulomb distortion effects on the fourth structure
function, at least around the first peak.  
From these results, we recommend that one can experimentally
extract the fourth structure function by choosing the ${\hat {\bf z}}$ axis
along the new modified momentum transfer ${\bf q}'(R)$.
Note that ${\theta}_{P}$ of the ejected proton in these plots is the polar angle
measured from the differently chosen ${\hat {\bf z}}$ axes.

Using a polarized incident electron beam and detecting the knocked
out proton out of the scattering plane, the fifth structure function
$R'_{LT'}$ can be extracted by measuring the up-down asymmetry of the
nuclear response.
We also choose the incident electron energy $E_{i}=500$ MeV, the
proton kinectic energy $T_{P}=100$ MeV, and the momentum transfer $q$
equal to the proton momentum $p$ for this case.

\begin{figure}[p]
\newbox\figj
\setbox\figj=\hbox{
\epsfysize=130mm
\epsfxsize=160mm
\epsffile{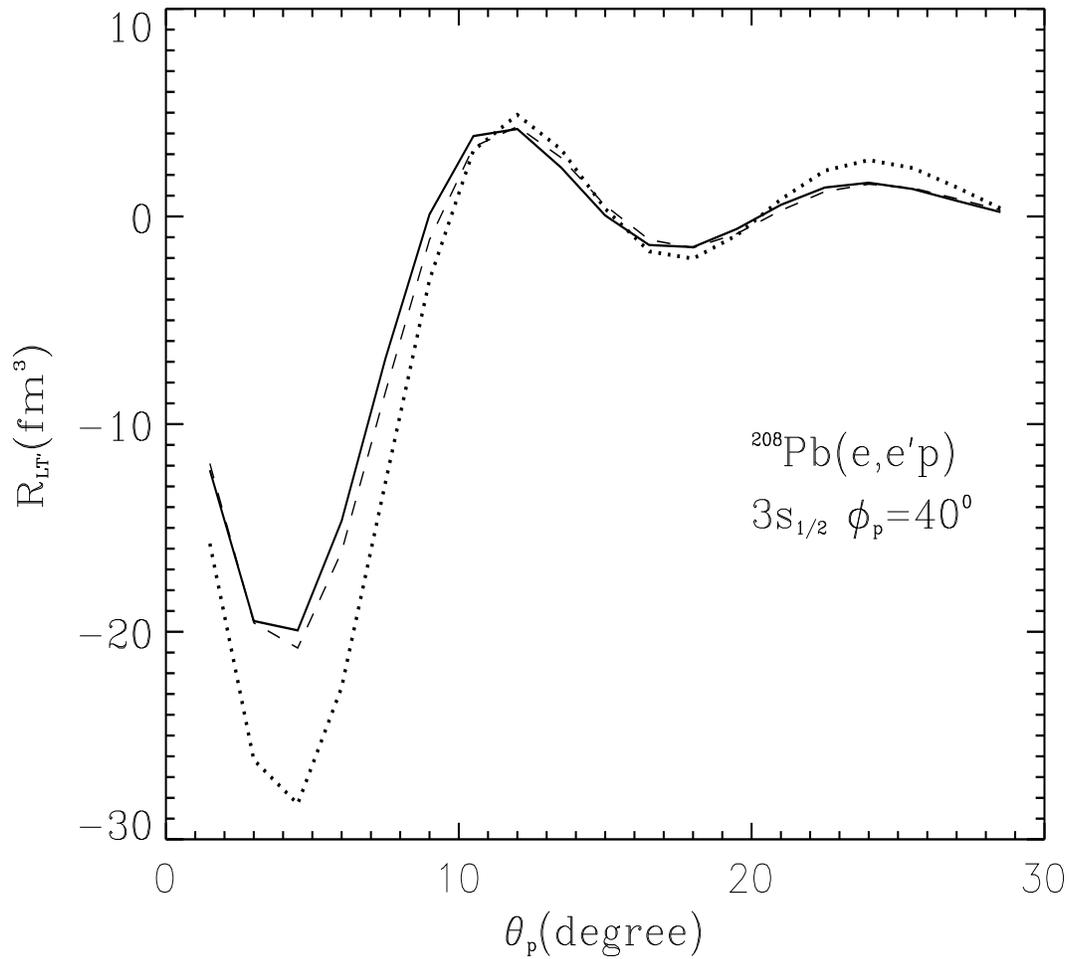}}
\noindent\hspace{0mm}\vspace{20mm}\box\figj
\caption{The fifth structure function for $^{208}Pb(e,e'p)$ from
the $3s_{1/2}$ shell as a function of the polar angle of the ejected
proton.  The dotted line is the fifth structure function for the
PWBA, while the solid  and the dashed lines are approximate DWBA results with
the ${\hat {\bf z}}$ axis along the asymptotic momentum ${\bf q}$ and along the
modified momentum transfer ${\bf q}'(R)$ respectively.}
\label{fives500}
\end{figure}

We first look at the Coulomb distortion effect on the measurement
of the fifth structure function from the $3s_{1/2}$ orbit of
$^{208}Pb$ at fixed proton azimuthal angle ${\phi}_{P}=40^{0}$ as
shown Fig.~(\ref{fives500}).
The direct fifth structure function again agrees with the apparent
structure function in PWBA calculation as expected.
The dotted line is the fifth structure function for the PWBA, and the
solid line is the fifth structure function with the ${\hat {\bf z}}$ axis
along the asymptotic momentum ${\bf q}$ and the dashed line is
for the case of the ${\hat {\bf z}}$ axis along the modified momentum
transfer ${\bf q}'(R)$.
We confirmed that the direct fifth structure function agrees with the
apparent structure function for the approximate DWBA calculation as 
expected from our earlier observation that the approximate 
 structure functions are symmetric under
the transformation $\phi_P \rightarrow -\phi_P$.  Thus when one calculates
$\sigma^U - \sigma^D$ all other terms cancel leaving only the
$R'_{LT'}$ contribution.  Note that unlike the case for $R'_{LT}$, 
changing the ${\hat {\bf z}}$ axis does not affect the shape and magnitude
and does not reduce the Coulomb distortion effect significantly.
Thus, 
one can experimentally extract a fifth structure function without
redefining the ${\hat {\bf z}}$ axis.  Of course, Coulomb distortion clearly affects the 
magnitude ot $R'_{LT'}$ as compared to the plane wave result (by approximately
30\% for $^{208}Pb$).

\begin{figure}[p]
\newbox\figk
\setbox\figk=\hbox{
\epsfysize=130mm
\epsfxsize=160mm
\epsffile{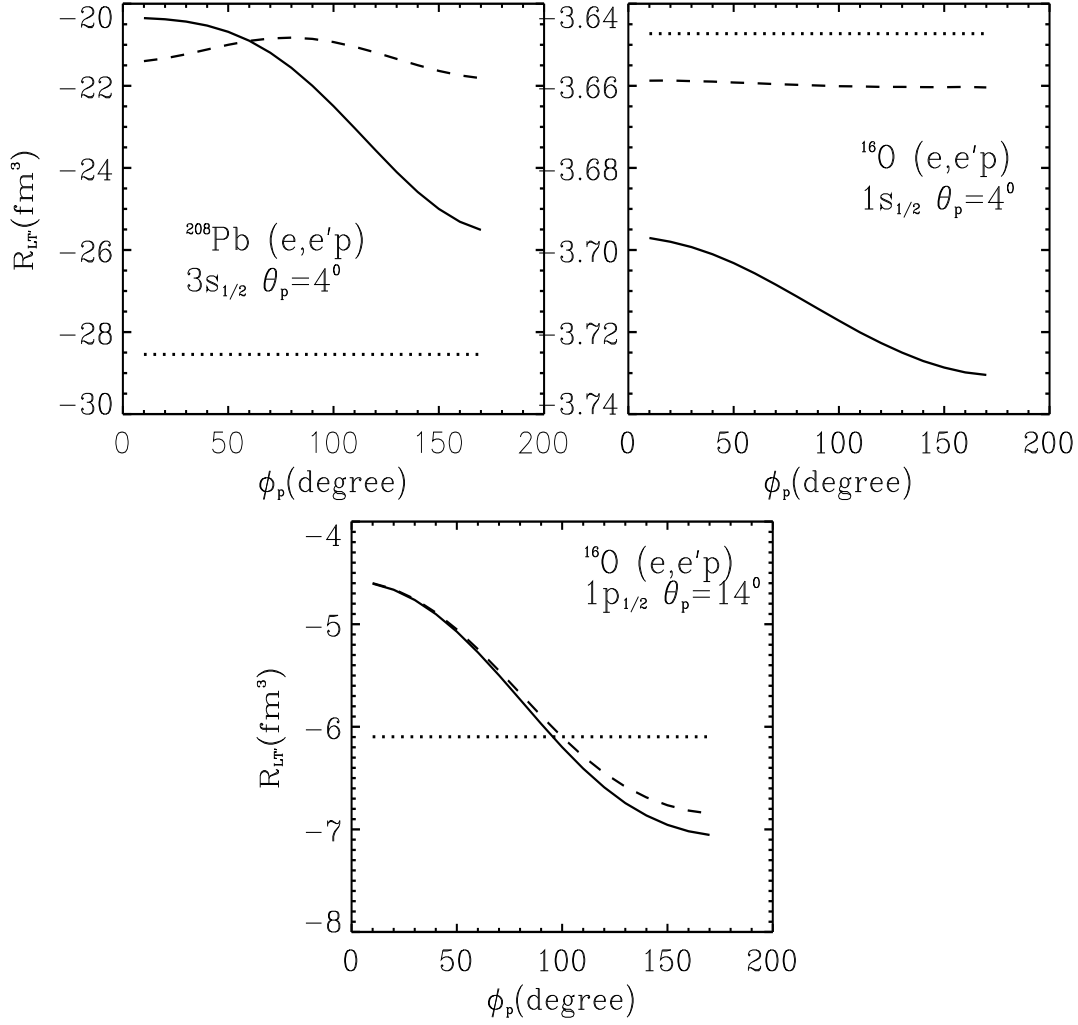}}
\noindent\hspace{0mm}\vspace{20mm}\box\figk
\caption{The fifth structure function as a function of the azimuthal 
angle ${\phi}_{P}$ for $^{208}Pb(e,e'p)$ from the $3s_{1/2}$
shell and for $^{16}O(e,e'p)$ from the $1s_{1/2}$ shell and the 
$1p_{1/2}$ shell.
The dotted line is the fifth structure function for PWBA, the solid
line is the fifth structure function with ${\hat {\bf z}}$ axis along the
asymptotic momentum ${\bf q}$, and the dashed line is for the case
of the ${\hat {\bf z}}$ axis along the momentum transfer ${\bf q}'(R)$ .}
\label{fif500}
\end{figure}

In Fig.~(\ref{fif500}), we show the fifth structure function as a function
of $\phi_P$ at
fixed polar proton angles, ${\theta}_{P}=4^{0}$, for the $3s_{1/2}$
orbit of $^{208}Pb$ and the $1s_{1/2}$ orbit of $^{16}O$, and
${\theta}_{P}=14^{0}$ for the $1p_{1/2}$ orbit of $^{16}O$.
These polar angles are the first peak position of the fifth structure
function for the $s_{1/2}$ orbit of $^{208}Pb$ and the $p_{1/2}$
orbit of $^{16}O$ respectively for these kinematics.
The dotted line is the PWBA result and the solid line and the dashed
line are the approximate DWBA results obtained by choosing the
${\hat {\bf z}}$ axis along the asymptotic momentum transfer ${\bf q}$ and
along the momentum transfer ${\bf q}'(R)$.
The approximate DWBA calculation for the $p_{1/2}$ orbit of
$^{16}O(e,e'p)$ in Fig.~(\ref{fif500}) reproduces the same shape as
the full DWBA calculation~\cite{jin94} for the fifth structure
function, but both differ in magnitude from the plane wave result.  In this
previous
paper~\cite{jin94} where the full DWBA calculation was applied to
a particular case, it was concluded that  extracting the
fifth structure function $R_{LT'}$
at ${\phi}_{P}=90^{0}$ or averaging over ${\phi}_{P}$ largely removed
the Coulomb distortion effects. Clearly that is not the case here. 
Therefore, it is not true in general that the Coulomb distortion effect can be
removed at ${\phi}_{P}=90^{0}$.
Furthermore, we again note that choosing a different
 ${\hat {\bf z}}$ axis does not help in
removing the Coulomb effect for the fifth structure function unlike
the case of the fourth structure function.
We also confirmed that the fifth structure function  extracted 
by using the
incident electron helicity dependence or the up-down asymmetry agree
exactly as expected from the $\phi_p \rightarrow - \phi_P$ symmetry
discussed earlier in the approximate Coulomb distorted form factors.

\section{Conclusions}

We have developed a plane-wave-like approximate solution to the Dirac equation
in the presence of the static Coulomb field of a nucleus which agrees rather
well with the exact partial wave solutions inside a sphere of approximately three
times the nuclear radius.  The limited spatial range of the approximation is not
a serious restriction for electron induced nuclear processes since the bound state
wavefunctions that enter any such process drop off exponentially outside the nuclear
radius R.  
Using this approximate wavefunction, along with a few additional approximations,
we also obtained an approximate DWBA potential valid for momentum transfers greater than
about $350$ MeV/c. This approximate potential has the same Dirac structure as
the plane wave  M${\o}$ller potential although it contains some spatially dependent
phase factors which destroy the azimuthal
spatial symmetry about the momentum transfer direction ${\bf q}$.  The basic ingredients in our approximate potential are the static
Coulomb potential of the target nucleus and the elastic scattering phase shifts for
the incoming and outgoing electron energies in this potential.  

We have compared wavefunctions, potentials and cross sections for the $(e,e'p)$
reaction on nuclei in the quasielastic region calculated with our approximation and 
previous approximations to the
exact partial wave results.  We find that the previous 
approximate results are in serious disagreement with the exact partial wave
results.   Our approximate results  are in good agreement 
with the full DWBA results, but have a number of 
advantages over the DWBA results.  The biggest  advantage  is that the 
plane-wave-like structure of the approximate Coulomb distorted
potential allows extraction of  structure functions  which are bi-linear products 
of transforms of the transition current components.  However, unlike the PWBA
analysis, these transforms are not just Fourier transforms, but contain
additional spatial dependence on the kinematics resulting from the static
Coulomb field of the target nucleus.   

In this paper, we investigated in some detail the extraction of
the so-called fourth structure function $R'_{LT}$ and fifth structure
function $R'_{LT'}$  from the full cross section.  We showed that this
is possible, particularly for kinematics where the structure functions 
are large.   However,  for the case of the fourth structure
function, $R'_{LT}$, the ${\hat {\bf z}}$ axis needs to be
redefined to lie along the momentum transfer defined at the nuclear surface
${\bf q}'(R)={\bf p}'_{i}(R)-{\bf p}'_{f}(R)$ to obtain agreement between
the directly calculated structure function and the extracted structure function.
For the fifth structure function $R'_{LT'}$, there is no need to redefine the
${\hat {\bf z}}$ axis.  

The other major advantage of our approximate treatment of Coulomb distortion over the 
full DWBA partial wave calculation is that it is straightforward to apply it to
higher energies.  The full DWBA calculation at higher electron energies requires 
more and more partial waves with increasingly forbidding amounts of computer time needed.
Using our approximate treatment, we do find it necessary to perform two additional 
numerical integrations over angular coordinates $\theta$ and
$\phi$ in the interaction matrix element as compared to a treatment that permits
a multipole decomposition.  However, these numerical integrations are
 not very time consuming and  in a sense are just a replacement for
summing over various intermediate angular momenta arising from angular momentum
recoupling of the various partial
wave expansions of the wavefunction and transition matrix elements.  
In our particular case, we have an
analytic result for the electromagnetic potential so that
the three dimensional integration is
very fast as compared to a full partial wave analysis and the
evaluation of thousands of radial matrix elements. 

In conclusion, our approximate treatment of Coulomb distortion for electron
induced nuclear processes involving continuum nucleons works quite well and
is particularly good for momentum transfers greater than 350 MeV/c.  It permits
the extraction of "structure functions"€which should prove of great use in analysing
$(e,e'p)$ experiments at the Thomas Jefferson National Accelerator Facility and other
laboratories. Of course,
it is an approximation and if one has very high precision data it may be necessary to
revert to the full DWBA calculation.

\begin{center}
ACKNOWLEDGMENTS
\end{center}

We thank the Ohio Supercomputer Center in Columbus for many hours
of Cray Y-MP time to develop this calculation and to perform the
necessary calculations. This work was supported in part by the U.S.
Department of Energy under Grant No. FG02-87ER40370.

\newpage

\end{document}